\documentclass[aps,prb,twocolumn,amsmath,amssymb,showpacs]{revtex4}

\usepackage{graphicx}
\usepackage{dcolumn}
\usepackage{bm}

\begin{document}

\title{Pairing symmetry in a two-orbital Hubbard model
on a square lattice}

\author{Katsunori Kubo}

\affiliation{
Advanced Science Research Center, Japan Atomic Energy Agency,
Tokai, Ibaraki 319-1195, Japan}

\received{26 December 2006;
revised manuscript received 21 February 2007;
published 14 June 2007}

\begin{abstract}
We investigate superconductivity in a two-orbital Hubbard model
on a square lattice by applying fluctuation exchange approximation.
In the present model, the symmetry of the two orbitals are assumed
to be that of an $s$ orbital.
Then, we find that
an $s$-wave spin-triplet orbital-antisymmetric state
and a $p$-wave spin-singlet orbital-antisymmetric state appear
when Hund's rule coupling is large.
These states are prohibited in a single-orbital model
within states with even frequency dependence,
but allowed for multi-orbital systems.
We also discuss pairing symmetry in other models
which are equivalent to the two-orbital Hubbard model
except for symmetry of orbitals.
Finally, we show that pairing states with a finite total momentum,
even without a magnetic field,
are possible in a system with two Fermi-surfaces.

\end{abstract}



\maketitle

\section{Introduction}

It has been recognized that orbital degree of freedom plays
important roles in determination of physical properties,
such as colossal magneto-resistance and complex ordered phases
of manganites,~\cite{Imada,Hotta}
and exotic magnetism in $f$-electron systems.~\cite{Hotta,Santini}
From experimental and theoretical studies of such systems,
it is revealed that magnetism in multi-orbital systems has a rich variety.
In recent years, effects of orbital degree of freedom 
on superconductivity have also been
discussed theoretically for several materials,~\cite{Takimoto2000, Takimoto2002, Takimoto2004, Mochizuki,Yanase, Mochizuki2, Yada, Kubo}
and it has been found that
orbital degree of freedom is important, e.g., for determination of
pairing symmetry in a system.
In particular, orbital degree of freedom probably plays a role
in triplet superconductivity of Sr$_2$RuO$_4$,~\cite{Maeno,Mackenzie}
in which Fermi surfaces are composed of $t_{2g}$ orbitals.

To understand the effects of orbital degree of freedom on superconductivity,
it is still important to gain a definite knowledge on those
in a relatively simple model such as a Hubbard model for two orbitals
with the same dispersion, although, at present,
it is difficult to find direct relevance of such a simple model
to actual materials.
For such a purpose,  superconductivity in two-orbital Hubbard models
has been studied by a mean-field theory~\cite{Klejnberg}
and by a dynamical mean-field theory.~\cite{Han,Sakai}
These studies have revealed that
an $s$-wave spin-triplet state, which satisfies the Pauli principle
by composing an orbital state of a pair antisymmetrically,
is a candidate for a ground state.
This fact is in sharp contrast to a single-orbital model
in which pairing states with even parity in the wavenumber space,
such as an $s$-wave state, should be spin-singlet ones
due to the Pauli principle within states with even frequency dependence.
Note that odd-frequency states are hard to be realized in ordinary cases.
In addition,
in a two-orbital system, it is also possible to realize
odd-parity spin-singlet states.
Thus, the variety of pairing states in a two-orbital model is much large.

However, in the above studies, possibility of superconductivity
other than the $s$-wave state is not considered.
In particular, within the standard dynamical mean-field theory,
i.e., in infinite spatial dimensions,
we cannot deal with spatial dependence of a pairing state.
Thus it is desirable to study superconductivity
in a multi-orbital model on a finite-dimensional lattice
and determine the most plausible candidates for pairing symmetry
of superconductivity.

In this paper, we investigate possible superconducting states
of a two-orbital Hubbard model on a square lattice
by applying fluctuation exchange (FLEX) approximation.
The FLEX approximation has been extended
to multi-orbital models.~\cite{Takimoto2004,Mochizuki,Mochizuki2,Yada,Kubo}
We classify superconducting states by spin states, orbital states,
and representations of tetragonal symmetry.
We also discuss pairing symmetry in other models
which are equivalent to the two-orbital Hubbard model.
In particular, we find that 
pairing states with a finite total momentum 
like the Fulde-Ferrell-Larkin-Ovchinnikov (FFLO) state,~\cite{Fulde,Larkin}
even without a magnetic field, are possible in a system with two Fermi-surfaces.

The organization of this paper is as follows.
In Sec.~\ref{sec:hamiltonian}, we introduce the two-orbital Hubbard model.
In Sec.~\ref{sec:formulation}, we explain the FLEX approximation
and categorize pairing symmetry of the model.
In Sec.~\ref{sec:results}, we show results
obtained with the FLEX approximation.
In Sec.~\ref{sec:other_models},
we discuss pairing symmetry in other models which are equivalent
to the two-orbital Hubbard model except for orbital symmetry.
We summarize the paper in Sec.~\ref{sec:summary}.

\section{Hamiltonian}
\label{sec:hamiltonian}

To investigate superconductivity in a multi-orbital system,
we consider a two-orbital Hubbard model given by
\begin{equation}
  \begin{split}
    H=&\sum_{\mathbf{k},\tau,\sigma}
    \epsilon_{\mathbf{k} \tau}
    c^{\dagger}_{\mathbf{k} \tau \sigma}c_{\mathbf{k} \tau \sigma}
    +U \sum_{i, \tau}
    n_{i \tau \uparrow} n_{i \tau \downarrow}\\
    &+U^{\prime} \sum_{i}
    n_{i 1} n_{i 2}
    + J \sum_{i,\sigma,\sigma^{\prime}}
    c^{\dagger}_{i 1 \sigma}
    c^{\dagger}_{i 2 \sigma^{\prime}}
    c_{i 1 \sigma^{\prime}}
    c_{i 2 \sigma}
    \\
    &+ J^{\prime}\sum_{i,\tau \ne \tau^{\prime}}
    c^{\dagger}_{i \tau \uparrow}
    c^{\dagger}_{i \tau \downarrow}
    c_{i \tau^{\prime} \downarrow}
    c_{i \tau^{\prime} \uparrow},
  \end{split}
  \label{eq:H}
\end{equation}
where $c_{i\tau\sigma}$ is the annihilation operator of
the electron at site $i$ with orbital $\tau$ ($=1$ or 2)
and spin $\sigma$ ($=\uparrow$ or $\downarrow$),
$c_{\mathbf{k}\tau\sigma}$ is the Fourier transform of it,
$n_{i \tau \sigma}=c^{\dagger}_{i \tau \sigma} c_{i \tau \sigma}$, and
$n_{i \tau}=\sum_{\sigma}n_{i \tau \sigma}$.
The coupling constants $U$, $U^{\prime}$, $J$, and $J^{\prime}$
denote the intra-orbital Coulomb, inter-orbital Coulomb, exchange,
and pair-hopping interactions, respectively.
In the followings, we use the relation
$U=U^{\prime}+J+J^{\prime}$,
which is satisfied in several orbital-degenerate models
such as a model for $p$-orbitals, a model for $e_g$ orbitals,
and a model for $t_{2g}$ orbitals.~\cite{Tang}
We also use the relation
$J=J^{\prime}$,
which holds if we can choose wavefunctions of orbitals real.~\cite{Tang}

Concerning the kinetic energy $\epsilon_{\mathbf{k}\tau}$,
we consider only a nearest-neighbor hopping
integral $t$ for both orbitals for simplicity,
and the kinetic energy is given by
$\epsilon_{\mathbf{k} 1}=\epsilon_{\mathbf{k} 2}=\epsilon_{\mathbf{k}}
=2t(\cos k_x+\cos k_y)$.
Here we have set the lattice constant unity.
We note that we assume that the symmetry of orbitals
is an $s$-orbital one, i.e.,
an orbital state does not change by a symmetry operation of a lattice,
such as inversion.
This assumption is crucial to determine pairing symmetry of superconductivity.

Here we note that the model Hamiltonian~\eqref{eq:H} can also describe
a system with different orbital-symmetry with a special condition.
For example, a model for doubly degenerate $p_x$ and $p_y$ orbitals
with the Slater-Koster integrals $(pp\sigma)=(pp\pi)=t$
is given by Eq.~\eqref{eq:H}
with $\epsilon_{\mathbf{k} 1}=\epsilon_{\mathbf{k} 2}=\epsilon_{\mathbf{k}}$.
Another model for $s$ orbitals described by Eq.~\eqref{eq:H}
with
$\epsilon_{\mathbf{k}1}=-\epsilon_{\mathbf{k}2}=\epsilon_{\mathbf{k}}$
is equivalent to the model with
$\epsilon_{\mathbf{k}1}=\epsilon_{\mathbf{k}2}=\epsilon_{\mathbf{k}}$,
if we change the phases of the wavefunctions for $\tau=2$ orbitals
by $\exp[i\mathbf{Q}\cdot \mathbf{r}_i]$ at each site $i$,
where $\mathbf{Q}=(\pi,\pi)$.
We also discuss pairing symmetry in such equivalent models
in Sec.~\ref{sec:other_models}.

\section{Formulation}
\label{sec:formulation}

In this section, we derive equations for response functions,
classify symmetry of superconductivity, and
derive a gap equation for the anomalous self-energy
for each symmetry.

\subsection{Green's function}
First, we derive equations for the Green's function in the normal phase.
In general, the Green's function is defined by
\begin{equation}
  G_{\tau_1 \sigma_1; \tau_2 \sigma_2}(\mathbf{k},\tau)
  =-
  \langle \text{T}_{\tau}
  c_{\mathbf{k} \tau_1 \sigma_1}(\tau)
  c^{\dagger}_{\mathbf{k} \tau_2 \sigma_2}
  \rangle,
\end{equation}
and the anomalous Green's functions are defined by
\begin{align}
  F_{\tau_1 \sigma_1; \tau_2 \sigma_2}(\mathbf{k},\tau)
  =-
  \langle \text{T}_{\tau}
  c_{\mathbf{k} \tau_1 \sigma_1}(\tau)
  c_{-\mathbf{k} \tau_2 \sigma_2}
  \rangle,
  \\
  F^{\dagger}_{\tau_1 \sigma_1; \tau_2 \sigma_2}(\mathbf{k},\tau)
  =-
  \langle \text{T}_{\tau}
  c^{\dagger}_{-\mathbf{k} \tau_1 \sigma_1}(\tau)
  c^{\dagger}_{\mathbf{k} \tau_2 \sigma_2}
  \rangle,
\end{align}
where
T$_{\tau}$ denotes the time-ordered product and
$\langle \cdots \rangle$ denotes the thermal average.
The Heisenberg representation for an operator $O$ is defined by
\begin{equation}
  O(\tau)
  =e^{\tau (H-\mu N_{\text{tot}})}
  O
  e^{-\tau (H-\mu N_{\text{tot}})},
\end{equation}
where $N_{\text{tot}}=\sum_{i, \tau}n_{i \tau}$ is the total number operator
of electrons and
$\mu$ is the chemical potential.
It is convenient to use the Fourier transformation
with respect to imaginary time given by
\begin{equation}
  O(i\epsilon_n)=\int^{\beta}_0 d\tau e^{i\epsilon_n \tau}O(\tau),
\end{equation}
where
$\beta=1/T$ with a temperature $T$ and
$\epsilon_n=(2n+1)\pi T$ is the Matsubara frequency for
fermions with an integer $n$.
Here we have set the Boltzmann constant unity.
Then, the Dyson-Gorkov equations are given by
\begin{equation}
  \begin{split}
    G_{\tau \sigma; \tau^{\prime} \sigma^{\prime}}(k)
    =&\delta_{\tau \tau^{\prime}}\delta_{\sigma \sigma^{\prime}}G^{(0)}(k)\\
    +\sum_{\tau^{\prime \prime}, \sigma^{\prime \prime}}
    [&G^{(0)}(k)
    \Sigma_{\tau \sigma; \tau^{\prime \prime} \sigma^{\prime \prime}}(k)
    G_{\tau^{\prime \prime} \sigma^{\prime \prime}; \tau^{\prime} \sigma^{\prime}}(k)\\
    +&G^{(0)}(k)
    \phi_{\tau \sigma; \tau^{\prime \prime} \sigma^{\prime \prime}}(k)
    F^{\dagger}_{\tau^{\prime \prime} \sigma^{\prime \prime}; \tau^{\prime} \sigma^{\prime}}(k)],
  \end{split}
\end{equation}
\begin{equation}
  \begin{split}
    F_{\tau \sigma; \tau^{\prime} \sigma^{\prime}}(k)\\
    =\sum_{\tau^{\prime \prime}, \sigma^{\prime \prime}}
    [&G^{(0)}(k)
    \Sigma_{\tau \sigma; \tau^{\prime \prime} \sigma^{\prime \prime}}(k)
    F_{\tau^{\prime \prime} \sigma^{\prime \prime}; \tau^{\prime} \sigma^{\prime}}(k)\\
    -&G^{(0)}(k)
    \phi_{\tau \sigma; \tau^{\prime \prime} \sigma^{\prime \prime}}(k)
    G_{\tau^{\prime} \sigma^{\prime}; \tau^{\prime \prime} \sigma^{\prime \prime}}(-k)],
  \end{split}
\end{equation}
where
$\Sigma_{\tau \sigma; \tau^{\prime} \sigma^{\prime}}(k)$ is the self-energy
and
$\phi_{\tau \sigma; \tau^{\prime} \sigma^{\prime}}(k)$ is the anomalous self-energy.
Here we have used the abbreviation $k=(\mathbf{k},i \epsilon_n)$.
The non-interacting Green's function is given by
\begin{equation}
  G^{(0)}(\mathbf{k},i\epsilon_n)=[i\epsilon_n-\epsilon_{\mathbf{k}}+\mu]^{-1}.
\end{equation}

In the normal state,
the Green's function and self-energy
do not depend on spin and orbital states, i.e.,
$G_{\tau \sigma; \tau^{\prime} \sigma^{\prime}}(k)
=\delta_{\tau \tau^{\prime}}\delta_{\sigma \sigma^{\prime}}G(k)$ and
$\Sigma_{\tau \sigma; \tau^{\prime} \sigma^{\prime}}(k)
=\delta_{\tau \tau^{\prime}}\delta_{\sigma \sigma^{\prime}}\Sigma(k)$,
and then the Dyson-Gorkov equation is given by
\begin{equation}
  G(k)=G^{(0)}(k)+G^{(0)}(k)\Sigma(k)G(k).
  \label{eq:DG_normal}
\end{equation}
The self-energy is given by
\begin{equation}
  \Sigma(k)=\frac{T}{N}\sum_{q}V(q)G(k-q),
\end{equation}
where
\begin{equation}
  V(q)=V^{\text{normal}}_{11;11}(q)+V^{\text{normal}}_{12;12}(q),
\end{equation}
in the FLEX approximation.
Here, $N$ is the number of lattice sites
and $q=(\mathbf{q},i\omega_m)$, where $\omega_m=2m\pi T$
is the Matsubara frequency for bosons with an integer $m$.
The matrix $V^{\text{normal}}(q)$ is given by
\begin{equation}
  \begin{split}
    V^{\text{normal}}(q)
    =\frac{3}{2}[U^{\text{s}} \chi^{\text{s}}(q) U^{\text{s}}
    -U^{\text{s}} \chi^{(0)}(q) U^{\text{s}}/2
    +U^{\text{s}}]
    \\
    +\frac{1}{2}[U^{\text{c}} \chi^{\text{c}}(q) U^{\text{c}}
    -U^{\text{c}} \chi^{(0)}(q) U^{\text{c}}/2
    -U^{\text{c}}].
  \end{split}
  \label{eq:V_normal}
\end{equation}
The matrix elements of $U^{\text{s}}$ and $U^{\text{c}}$ are given by
\begin{align}
  &U^{\text{s}}_{11;11}=U^{\text{s}}_{22;22}
  =U^{\text{c}}_{11;11}=U^{\text{c}}_{22;22}=U,\\
  &U^{\text{s}}_{11;22}=U^{\text{s}}_{22;11}=J,\\
  &U^{\text{c}}_{11;22}=U^{\text{c}}_{22;11}=2U^{\prime}-J,\\
  &U^{\text{s}}_{12;12}=U^{\text{s}}_{21;21}=U^{\prime},\\
  &U^{\text{c}}_{12;12}=U^{\text{c}}_{21;21}=-U^{\prime}+2J,\\
  &U^{\text{s}}_{12;21}=U^{\text{s}}_{21;12}
  =U^{\text{c}}_{12;21}=U^{\text{c}}_{21;12}=J^{\prime},
\end{align}
and zero for the other elements of these matrices.
The matrices for susceptibilities $\chi^{\text{s}}(q)$ for the spin part
and $\chi^{\text{c}}(q)$ for the charge part
are given by
\begin{align}
  \chi^{\text{s}}(q)=\chi^{(0)}(q)[1-U^{\text{s}} \chi^{(0)}(q)]^{-1},
  \label{eq:chi_s}
  \\
  \chi^{\text{c}}(q)=\chi^{(0)}(q)[1+U^{\text{c}} \chi^{(0)}(q)]^{-1},
\end{align}
where
\begin{equation}
  \chi^{(0)}(q)=-\frac{T}{N}\sum_{k}G(k+q)G(k).
  \label{eq:chi_0}
\end{equation}
We solve Eqs.~\eqref{eq:DG_normal}-\eqref{eq:V_normal}
and \eqref{eq:chi_s}-\eqref{eq:chi_0} self-consistently.

\subsection{Response functions}

By using obtained $\chi^{\text{s}}(q)$ and $\chi^{\text{c}}(q)$,
we can calculate response functions.
The response function corresponding to an operator $O^{\text{A}}_i$
is given by
\begin{equation}
    \chi^{\text{A}}(\mathbf{q},i\omega_m)
    =
    \sum_{i} \int^{\beta}_0 d\tau
    e^{-i \mathbf{q} \cdot \mathbf{r}_i+i \omega_m \tau}
    \langle \text{T}_{\tau}
    O^{\text{A}}_{i}(\tau)
    O^{\text{A}}_{\text{o}}
    \rangle,
\end{equation}
where o denotes the origin.
An operator $O^{\text{A}}_{i}$ in the second-quantized form is given by
\begin{equation}
  O^{\text{A}}_{i}
  =\sum_{\tau, \tau^{\prime}, \sigma, \sigma^{\prime}}
  c^{\dagger}_{i \tau \sigma}
  O^{\text{A}}_{\tau \sigma; \tau^{\prime} \sigma^{\prime}}
  c_{i \tau^{\prime} \sigma^{\prime}}.
\end{equation}
The matrix elements $O^{\text{A}}_{\tau \sigma; \tau^{\prime} \sigma^{\prime}}$
are given by
\begin{align}
  O^{\text{charge}}_{\tau \sigma; \tau^{\prime} \sigma^{\prime}}
  &=\delta_{\tau \tau^{\prime}}\delta_{\sigma \sigma^{\prime}},\\
  O^{\sigma^{\nu}}_{\tau \sigma; \tau^{\prime} \sigma^{\prime}}
  &=\delta_{\tau \tau^{\prime}}\hat{\sigma}^{\nu}_{\sigma \sigma^{\prime}},\\
  O^{\tau^{\nu}}_{\tau \sigma; \tau^{\prime} \sigma^{\prime}}
  &=\hat{\sigma}^{\nu}_{\tau \tau^{\prime}}\delta_{\sigma \sigma^{\prime}},\\
  O^{\tau^{\nu} \sigma^{\nu^{\prime}}}_{\tau \sigma; \tau^{\prime} \sigma^{\prime}}
  &=\hat{\sigma}^{\nu}_{\tau \tau^{\prime}}\hat{\sigma}^{\nu^{\prime}}_{\sigma \sigma^{\prime}},
\end{align}
for charge, spin, orbital, and
spin-orbital coupled operators, respectively,
where $\hat{\sigma}^{\nu}$ is the Pauli matrix for
$\nu$ ($=x$, $y$, or $z$) component.
Due to the rotational symmetry in the spin space,
the relations
$\chi^{\sigma^x}(q)=\chi^{\sigma^y}(q)=\chi^{\sigma^z}(q)$ and
$\chi^{\tau^{\nu}\sigma^x}(q)
=\chi^{\tau^{\nu}\sigma^y}(q)
=\chi^{\tau^{\nu}\sigma^z}(q)$ hold.
In addition, the present model has rotational symmetry
in the $\tau^z$-$\tau^x$ plane, and the relations
$\chi^{\tau^x}(q)=\chi^{\tau^z}(q)$ and
$\chi^{\tau^x \sigma^z}(q)=\chi^{\tau^z \sigma^z}(q)$
also hold.
We note that there is additional symmetry for $J=0$.
For $J=0$,
the model is invariant under the transformation
$c_{i 2 \downarrow} \rightarrow -c_{i 2 \downarrow}$,
which transforms $O^{\tau^y}_i$ to $O^{\tau^y \sigma^z}_i$.
In addition, for $J=0$,
the orbital space also has the full rotational symmetry
and is equivalent to the spin space,
and thus all the above response functions are the same
except for $\chi^{\text{charge}}(q)$.

The response functions in the FLEX approximation are given by
\begin{align}
  \chi^{\text{charge}  }(q)=4[\chi^{c}_{11;11}(q)+\chi^{c}_{11;22}(q)],\\
  \chi^{\sigma^z       }(q)=4[\chi^{s}_{11;11}(q)+\chi^{s}_{11;22}(q)],
  \label{eq:chi_sigma}\\
  \chi^{\tau^x         }(q)=4[\chi^{c}_{12;12}(q)+\chi^{c}_{12;21}(q)],\\
  \chi^{\tau^y         }(q)=4[\chi^{c}_{12;12}(q)-\chi^{c}_{12;21}(q)],\\
  \chi^{\tau^z         }(q)=4[\chi^{c}_{11;11}(q)-\chi^{c}_{11;22}(q)],\\
  \chi^{\tau^x \sigma^z}(q)=4[\chi^{s}_{12;12}(q)+\chi^{s}_{12;21}(q)],\\
  \chi^{\tau^y \sigma^z}(q)=4[\chi^{s}_{12;12}(q)-\chi^{s}_{12;21}(q)],\\
  \chi^{\tau^z \sigma^z}(q)=4[\chi^{s}_{11;11}(q)-\chi^{s}_{11;22}(q)],
\end{align}
where we have used trivial relations such as
$\chi^{c}_{11;11}(q)=\chi^{c}_{22;22}(q)$.
Within the FLEX approximation, we can show the relation
$\chi^{\tau^y}(q)=\chi^{\tau^y \sigma^z}(q)$ even for $J \ne 0$,
while it does not hold in general.

\subsection{Gap equation}

Now, we derive a gap equation for superconductivity.
First, we categorize the anomalous self-energy by symmetry.
The anomalous self-energy for a spin-singlet state is given by
\begin{equation}
  \phi^{\text{singlet}}_{\tau \tau^{\prime}}(k)
  =\frac{1}{2}[\phi_{\tau \uparrow; \tau^{\prime} \downarrow}(k)
  -\phi_{\tau \downarrow; \tau^{\prime} \uparrow}(k)].
\end{equation}
The anomalous self-energy for a spin-triplet state is given by
\begin{equation}
  \phi^{\text{triplet}}_{\tau \tau^{\prime}}(k)
  =\frac{1}{2}[\phi_{\tau \uparrow; \tau^{\prime} \downarrow}(k)
  +\phi_{\tau \downarrow; \tau^{\prime} \uparrow}(k)].
\end{equation}
Due to the rotational symmetry in the spin space,
the spin-triplet states with
$\phi^{\text{triplet}}_{\tau \tau^{\prime}}(k)$,
with $\phi_{\tau \uparrow; \tau^{\prime} \uparrow}(k)$,
and with $\phi_{\tau \downarrow; \tau^{\prime} \downarrow}(k)$
are degenerate.
We can categorize superconducting states further by orbital symmetry.
For an orbital-parallel-antisymmetric state (orbital-$\tau^x$ state),
the anomalous self-energy is defined by
\begin{equation}
  \begin{split}
    \phi^{\xi}_{x}(k)
    &=-\frac{i}{2}\sum_{\tau,\tau^{\prime},\tau^{\prime \prime}}
    \sigma^y_{\tau \tau^{\prime}}
    \sigma^x_{\tau^{\prime} \tau^{\prime \prime}}
    \phi^{\xi}_{\tau^{\prime \prime} \tau}(k)
    \\
    &=-\frac{1}{2}[\phi^{\xi}_{11}(k)-\phi^{\xi}_{22}(k)].
  \end{split}
\end{equation}
For an orbital-parallel-symmetric state (orbital-$\tau^y$ state),
the anomalous self-energy is defined by
\begin{equation}
  \begin{split}
    \phi^{\xi}_{y}(k)
    &=-\frac{i}{2}\sum_{\tau,\tau^{\prime},\tau^{\prime \prime}}
    \sigma^y_{\tau \tau^{\prime}}
    \sigma^y_{\tau^{\prime} \tau^{\prime \prime}}
    \phi^{\xi}_{\tau^{\prime \prime} \tau}(k)
    \\
    &=-\frac{i}{2}[\phi^{\xi}_{11}(k)+\phi^{\xi}_{22}(k)].
  \end{split}
\end{equation}
For an orbital-antiparallel-symmetric state (orbital-$\tau^z$ state),
the anomalous self-energy is defined by
\begin{equation}
  \begin{split}
    \phi^{\xi}_{z}(k)
    &=-\frac{i}{2}\sum_{\tau,\tau^{\prime},\tau^{\prime \prime}}
    \sigma^y_{\tau \tau^{\prime}}
    \sigma^z_{\tau^{\prime} \tau^{\prime \prime}}
    \phi^{\xi}_{\tau^{\prime \prime} \tau}(k)
    \\
    &=\frac{1}{2}[\phi^{\xi}_{12}(k)+\phi^{\xi}_{21}(k)].
  \end{split}
\end{equation}
For an orbital-antiparallel-antisymmetric state (orbital-$\tau^0$ state),
the anomalous self-energy is defined by
\begin{equation}
  \begin{split}
    \phi^{\xi}_{0}(k)
    &=-\frac{i}{2}\sum_{\tau,\tau^{\prime},\tau^{\prime \prime}}
    \sigma^y_{\tau \tau^{\prime}}
    \delta_{\tau^{\prime} \tau^{\prime \prime}}
    \phi^{\xi}_{\tau^{\prime \prime} \tau}(k)
    \\
    &=\frac{1}{2}[\phi^{\xi}_{12}(k)-\phi^{\xi}_{21}(k)].
  \end{split}
\end{equation}
An orbital-$\tau^{\nu}$ state with $\nu=x$, $y$, or $z$
is a state with a $d$-vector for the orbital space
parallel to the $\nu$ axis,
while an orbital-$\tau^0$ state is an orbital-singlet state.
The anomalous self-energy has the following symmetry:
\begin{equation}
  \phi_{\tau \sigma; \tau^{\prime} \sigma^{\prime}}(-k)
  =-\phi_{\tau^{\prime} \sigma^{\prime}; \tau \sigma}(k).
\end{equation}
Then the following relations are obtained:
\begin{align}
  \phi^{\text{singlet}}_{\nu}(-k)
  &=\phi^{\text{singlet}}_{\nu}(k),\\
  \phi^{\text{triplet}}_{\nu}(-k)
  &=-\phi^{\text{triplet}}_{\nu}(k),\\
  \phi^{\text{singlet}}_{0}(-k)
  &=-\phi^{\text{singlet}}_{0}(k),
  \label{eq:singlet_AA}
  \\
  \phi^{\text{triplet}}_{0}(-k)
  &=\phi^{\text{triplet}}_{0}(k),
  \label{eq:triplet_AA}
\end{align}
where $\nu=x$, $y$, or $z$.

The linearized gap equation for the anomalous self-energy is expressed as
\begin{equation}
  \begin{split}
    \lambda(\Gamma,\xi,\tilde{\nu})
    \phi^{\xi}_{\tilde{\nu}}(k)
    &=\frac{T}{N}\sum_{k^{\prime}}
    V^{\xi}_{\tilde{\nu}}(k-k^{\prime})F^{\xi}_{\tilde{\nu}}(k^{\prime}),\\
    =&-\frac{T}{N}\sum_{k^{\prime}}
    V^{\xi}_{\tilde{\nu}}(k-k^{\prime})|G(k^{\prime})|^2
    \phi^{\xi}_{\tilde{\nu}}(k^{\prime}),
  \end{split}
  \label{eq:gap_eq}
\end{equation}
with $\lambda(\Gamma,\xi,\tilde{\nu})=1$,
where $\Gamma$ denotes a representation of tetragonal symmetry C$_{4v}$
which $\phi^{\xi}_{\tilde{\nu}}(k)$ obeys,
$\tilde{\nu}=x$, $y$, $z$, or 0, and
$F^{\xi}_{\tilde{\nu}}(k)$ is defined by the same way
as $\phi^{\xi}_{\tilde{\nu}}(k)$.
Thus, the superconducting transition temperature is given by
the temperature where an eigenvalue $\lambda(\Gamma,\xi,\tilde{\nu})$
of Eq.~\eqref{eq:gap_eq} becomes unity.
The effective pairing interactions $V^{\xi}_{\tilde{\nu}}(q)$ are written as
\begin{align}
  V^{\xi}_{x}(q)=V^{\xi}_{11;11}(q)-V^{\xi}_{12;21}(q),
  \label{eq:V_PA}\\
  V^{\xi}_{y}(q)=V^{\xi}_{11;11}(q)+V^{\xi}_{12;21}(q),
  \label{eq:V_PS}\\
  V^{\xi}_{z}(q)=V^{\xi}_{11;22}(q)+V^{\xi}_{12;12}(q),
  \label{eq:V_AS}\\
  V^{\xi}_{0}(q)=V^{\xi}_{11;22}(q)-V^{\xi}_{12;12}(q),
  \label{eq:V_AA}
\end{align}
where
\begin{equation}
  \begin{split}
    V^{\text{singlet}}(q)
    =&\frac{3}{2}[U^{\text{s}} \chi^{\text{s}}(q) U^{\text{s}}
    +U^{\text{s}}/2]
    \\
    -&\frac{1}{2}[U^{\text{c}} \chi^{\text{c}}(q) U^{\text{c}}
    -U^{\text{c}}/2],\\
  \end{split}
\end{equation}
\begin{equation}
  \begin{split}
    V^{\text{triplet}}(q)
    =&-\frac{1}{2}[U^{\text{s}} \chi^{\text{s}}(q) U^{\text{s}}
    +U^{\text{s}}/2]
    \\
    &-\frac{1}{2}[U^{\text{c}} \chi^{\text{c}}(q) U^{\text{c}}
    -U^{\text{c}}/2].
  \end{split}
  \label{eq:V_triplet}
\end{equation}

Due to the rotational symmetry of the model
in the $\tau^z$-$\tau^x$ plane,
the orbital-$\tau^z$ state and the orbital-$\tau^x$ state are degenerate.
In addition, for $J=0$,
the orbital space has full rotational symmetry, and
orbital-$\tau^x$, -$\tau^y$, and -$\tau^z$ states are degenerate, that is,
they are orbital-triplet states.
Moreover, in this case,
the spin space and the orbital space are equivalent,
and thus,
a spin-singlet orbital-triplet state and
a spin-triplet orbital-singlet state are degenerate.
Note also that,
by changing phases of wavefunctions,
we can show that
a spin-singlet orbital-singlet state and
a spin-triplet orbital-triplet state are degenerate for $J=0$.

Before presenting calculated results,
here we briefly discuss possible candidates
for pairing symmetry of superconductivity.
For $s$-wave pairing,
a spin-triplet state is favorable
since a solution $\phi^{\xi}_{\tilde{\nu}}(k)$ does not have to change its sign
in the $\mathbf{k}$ space
for a spin-triplet state due to the sign in Eq.~\eqref{eq:V_triplet}
when fluctuations $\chi^{\text{s}}(q)$ and/or $\chi^{\text{c}}(q)$ are large.
In single-orbital models,
such an $s$-wave spin-triplet state is odd in frequency,
which hardly appears in ordinary cases.
However, in the present model,
an $s$-wave spin-triplet state with even frequency dependence
is allowed for an orbital-$\tau^0$ state [see Eq.~\eqref{eq:triplet_AA}].
For a $p$-wave state,
a spin-triplet state is unfavorable,
since $\phi^{\xi}_{\tilde{\nu}}(k)$ should change its sign in the $\mathbf{k}$ space
for a $p$-wave state.
Thus, a spin-singlet state is favorable for $p$-wave pairing,
which is allowed in the present model for an orbital-$\tau^0$ state
with even frequency dependence [see Eq.~\eqref{eq:singlet_AA}].
For a $d_{x^2-y^2}$ state,
a spin-singlet state is favorable
because of similar logic for $p$-wave pairing,
while such a state is an orbital-$\tau^x$, -$\tau^y$, or -$\tau^z$ state.

\section{Results}
\label{sec:results}

In this section,
we show results for a $64 \times 64$ lattice.
In the calculation, we use 2048 Matsubara frequencies.
In this study, we fix the value of
the intra-orbital Coulomb interaction $U=6t$
and vary $J$ ($=J^{\prime}$).
Then the inter-orbital Coulomb interaction is given by
$U^{\prime}=U-2J$.
In the followings, we discuss superconducting states
with even frequency dependence only,
since we find that eigenvalues $\lambda(\Gamma,\xi,\tilde{\nu})$
for odd-frequency states are small.

\begin{figure}[t]
  \includegraphics[width=1\linewidth]{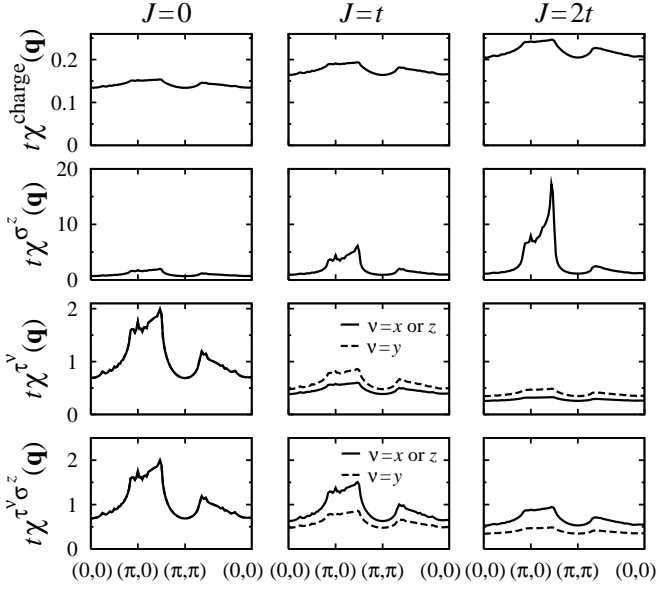}
  \caption{\label{figure:chi}
    $\mathbf{q}$ dependence of
    the susceptibilities for $J=0$, $t$, and $2t$
    at $T=0.005t$, $n=1$, and $U=6t$.
  }
\end{figure}
Figure~\ref{figure:chi} shows static susceptibilities
$\chi^{\text{A}}(\mathbf{q})=\chi^{\text{A}}(\mathbf{q},i\omega_m=0)$
for $J=0$, $t$, and $2t$ at $T=0.005t$
and the electron number $n=\langle N_{\text{tot}} \rangle /N=1$ per site.
Among the susceptibilities,
the spin susceptibility $\chi^{\sigma^z}(\mathbf{q})$ is strongly enhanced by
increasing $J$, that is,
such magnetic fluctuations are enhanced by the Hund's rule coupling.
On the other hand,
$\chi^{\tau^{\nu}}(\mathbf{q})$ and $\chi^{\tau^{\nu} \sigma^z}(\mathbf{q})$,
which include orbital fluctuations,
are suppressed by the Hund's rule coupling.
The charge susceptibility $\chi^{\text{charge}}(\mathbf{q})$
is enhanced a little by the Hund's rule coupling,
but it is still small.
Thus, among various fluctuations,
the spin fluctuations for a large $J$ are important
in the present model.

\begin{figure}[t]
  \includegraphics[width=7.5cm]{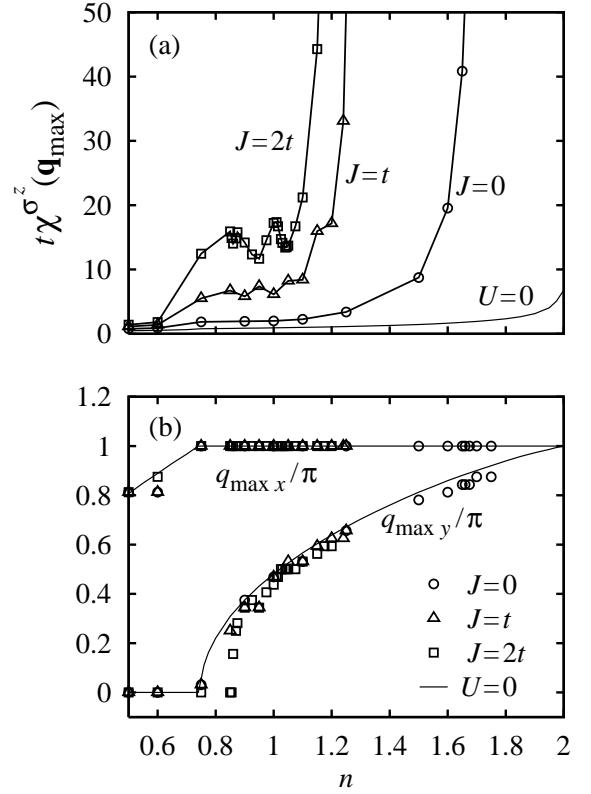}
  \caption{\label{figure:max_q}
    (a) Spin susceptibilities
    $\chi^{\sigma^z}(\mathbf{q})$
    at $\mathbf{q}=\mathbf{q}_{\text{max}}$
    and (b) $\mathbf{q}_{\text{max}}$
    for $J=0$, $t$, and $2t$
    as functions of $n$
    at $T=0.005t$ and $U=6t$.
    The thin solid lines represent those for $U=0$.
  }
\end{figure}
In Fig.~\ref{figure:max_q}(a),
we show $n$ dependence of $\chi^{\sigma^z}(\mathbf{q}_{\text{max}})$
for $J=0$, $t$, and $2t$ at $T=0.005t$,
where $\mathbf{q}_{\text{max}}$ is defined as the wavevector
at which $\chi^{\sigma^z}(\mathbf{q})$ becomes the largest.
For comparison, we also show $\chi^{\sigma^z}(\mathbf{q}_{\text{max}})$
for the non-interacting system.
The spin susceptibility is enhanced by the Coulomb interaction,
and is further increased by the Hund's rule coupling
as is already shown in Fig.~\ref{figure:chi} for $n=1$.
However, $\mathbf{q}_{\text{max}}$ does not depend so much on
the Coulomb interaction and the Hund's rule coupling
as shown in Fig.~\ref{figure:max_q}(b).
This fact indicates that
the characteristic wavevector within the FLEX approximation
is almost determined by the property of the non-interacting system, i.e.,
the Fermi-surface structure.
As shown in Fig.~\ref{figure:max_q}(b),
the characteristic wavevector is $\mathbf{q}_{\text{max}} \simeq (\pi,\pi)$
around $n=2$,
and becomes $\mathbf{q}_{\text{max}} \simeq (\pi,0)$
by decreasing $n$ to $n \simeq 0.8$.
Thus, it may be expected that
a superconducting state with $d_{x^2-y^2}$ symmetry appears
for a large $n$
and 
a $p$-wave superconducting state appears for a small $n$.
We also expect an occurrence of an $s$-wave state
if the enhanced spin-fluctuations
in a large part of the $\mathbf{q}$ space are available.

Concerning the orbital state,
only the orbital-$\tau^0$ states are possible for
the $s$-wave spin-triplet and $p$-wave spin-singlet states
as is discussed in the previous section.
For $d_{x^2-y^2}$-wave spin-singlet pairing,
there are three possible orbital states, $\tau^x$, $\tau^y$, and $\tau^z$.
By increasing the Hund's rule coupling, spin fluctuations become dominant,
and superconductivity is mainly determined by
$\chi^{\text{s}}_{11;11}(q)$ or $\chi^{\text{s}}_{11;22}(q)$
[see Eq.~\eqref{eq:chi_sigma}].
For $J=t$ and $2t$,
from the calculated results shown in Fig.~\ref{figure:chi},
$\chi^{\tau^y \sigma^z}(\mathbf{q})<\chi^{\tau^x \sigma^z}(\mathbf{q})$ and
$\chi^{\tau^y \sigma^z}(\mathbf{q})<\chi^{\tau^z \sigma^z}(\mathbf{q})$,
we obtain
\begin{align}
   \chi^{\text{s}}_{11;11}(\mathbf{q})
  -\chi^{\text{s}}_{12;21}(\mathbf{q})
  &<
   \chi^{\text{s}}_{11;11}(\mathbf{q})
  +\chi^{\text{s}}_{12;21}(\mathbf{q}),\\
   \chi^{\text{s}}_{11;22}(\mathbf{q})
  +\chi^{\text{s}}_{12;12}(\mathbf{q})
  &<
   \chi^{\text{s}}_{11;11}(\mathbf{q})
  +\chi^{\text{s}}_{12;21}(\mathbf{q}),
\end{align}
respectively.
Thus, from Eqs.~\eqref{eq:V_PA}-\eqref{eq:V_AS},
we expect that an orbital-$\tau^y$ state is the most favorable state
among the $d_{x^2-y^2}$-wave spin-singlet states.
However, the difference among $V^{\text{singlet}}_{\nu}(q)$
($\nu=x$, $y$, and $z$) would be small,
since only
$\chi^{\text{s}}_{11;11}(\mathbf{q})$ and
$\chi^{\text{s}}_{11;22}(\mathbf{q})$ are large and
$\chi^{\sigma^z}(\mathbf{q})\gg\chi^{\tau^z \sigma^z}(\mathbf{q})>0$
means
$\chi^{\text{s}}_{11;11}(\mathbf{q}) \simeq \chi^{\text{s}}_{11;22}(\mathbf{q})$.

\begin{figure}[t]
  \includegraphics[width=1\linewidth]{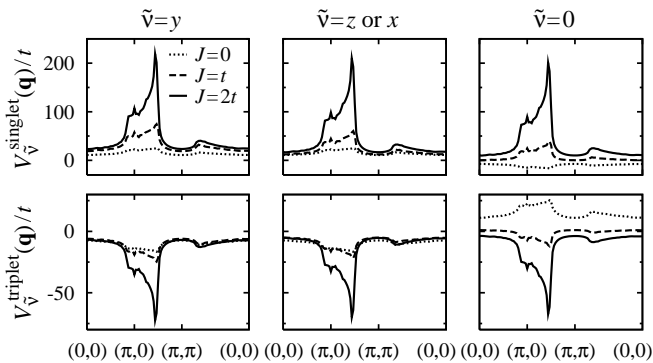}
  \caption{\label{figure:V_super}
    $\mathbf{q}$ dependence of
    the effective pairing interactions $V^{\xi}_{\tilde{\nu}}(\mathbf{q})$
    for $J=0$, $t$, and $2t$
    at $T=0.005t$, $n=1$, and $U=6t$.
  }
\end{figure}
Figure~\ref{figure:V_super} shows $\mathbf{q}$ dependence of
the effective pairing interactions
$V^{\xi}_{\tilde{\nu}}(\mathbf{q})=V^{\xi}_{\tilde{\nu}}(\mathbf{q},i\omega_m=0)$
at zero frequency for $J=0$, $t$, and $2t$ at $T=0.005t$ and $n=1$.
The magnitude of the effective interactions are increased
by the Hund's rule coupling.
As is discussed above,
$V^{\text{singlet}}_{y}(\mathbf{q})$
is slightly larger than
$V^{\text{singlet}}_{z}(\mathbf{q})$
and
$V^{\text{singlet}}_{x}(\mathbf{q})$.
For $J=0$,
$V^{\text{triplet}}_{0}(\mathbf{q})$
is repulsive for $s$-wave pairing,
but becomes attractive for $J=2t$.
For $J=0$, all the susceptibilities are the same except for the charge one,
and we obtain
\begin{equation}
  \begin{split}
    \chi^{\text{s}}_{11;22}(\mathbf{q})
    -\chi^{\text{s}}_{12;12}(\mathbf{q})
    &=-\chi^{\text{s}}_{12;12}(\mathbf{q})\\
    &=-\chi^{\tau^x}(\mathbf{q})/4<0.
  \end{split}
\end{equation}
From the inequalities
$\chi^{\text{charge}}(\mathbf{q})<\chi^{\tau^z}(\mathbf{q})$ and
$0<\chi^{\tau^x}(\mathbf{q})=4\chi^{\text{c}}_{12;12}(\mathbf{q})$,
we obtain
$\chi^{\text{c}}_{11;22}(\mathbf{q})<0<\chi^{\text{c}}_{12;12}(\mathbf{q})$,
that is,
\begin{equation}
  \chi^{\text{c}}_{11;22}(\mathbf{q})-\chi^{\text{c}}_{12;12}(\mathbf{q})<0,
\end{equation}
for $J=0$.
Thus,
$V^{\text{triplet}}_{0}(\mathbf{q})$ is repulsive
for $J=0$ [see Eqs.~\eqref{eq:V_AA} and \eqref{eq:V_triplet}].
For $J=2t$, $\chi^{\text{s}}_{11;22}(\mathbf{q})$ is large,
and the spin-triplet orbital-$\tau^0$ pairing interaction becomes
attractive.

\begin{figure}[t]
  \includegraphics[width=7cm]{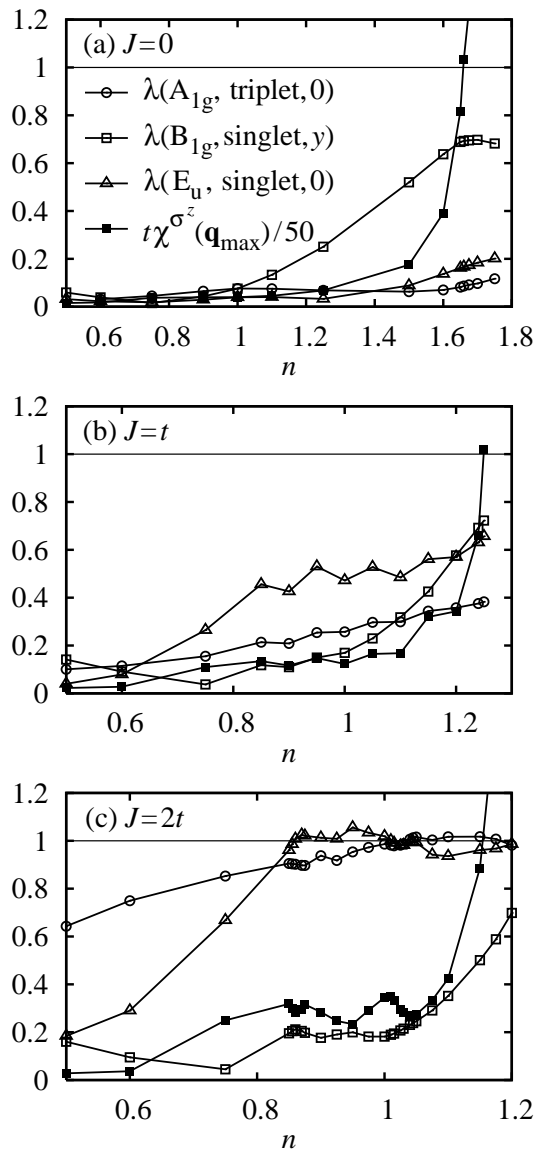}
  \caption{\label{figure:lambda_chi}
    Eigenvalues
    $\lambda(\text{A}_{1g},\text{triplet},0)$,
    $\lambda(\text{B}_{1g},\text{singlet},y)$, and
    $\lambda(\text{E}_{ u},\text{singlet},0)$, and
    the spin susceptibilities
    $\chi^{\sigma^z}(\mathbf{q}_{\text{max}})$
    as functions of $n$
    for (a) $J=0$,  (b) $J=t$, and  (c) $J=2t$
    at $T=0.005t$ and $U=6t$.
  }
\end{figure}
Among all the possible superconducting states,
we find that only three states,
the $s$-wave (A$_{1g}$ symmetry) spin-triplet orbital-$\tau^0$ state,
the $p$-wave (E$_{ u}$ symmetry) spin-singlet orbital-$\tau^0$ state, and
the $d_{x^2-y^2}$-wave (B$_{1g}$ symmetry) spin-singlet orbital-$\tau^y$ state,
have the largest eigenvalues of the gap equation Eq.~\eqref{eq:gap_eq}
with $\lambda(\Gamma,\xi,\tilde{\nu}) > 0.5$
for some parameter sets we have used.
Figures~\ref{figure:lambda_chi}(a)--\ref{figure:lambda_chi}(c)
show $n$ dependence of eigenvalues for these states
and $\chi^{\sigma^z}(\mathbf{q}_{\text{max}})$ at $T=0.005t$
for $J=0$, $t$, and $2t$, respectively.
The eigenvalues are enhanced by the Hund's rule coupling
as the spin susceptibility.
In particular, for $J=2t$,
$\lambda(\text{A}_{1g},\text{triplet},0)$
and
$\lambda(\text{E}_{u},\text{singlet},0)$
become larger than unity,
that is, superconducting transitions occur
for these symmetry with transition temperatures higher than $0.005t$.
The $p$-wave superconductivity appears around $n=0.9$,
where the characteristic wavevector is $\mathbf{q}_{\text{max}} \simeq (\pi,0)$
[see Fig.~\ref{figure:max_q}(b)].
For the $s$-wave state,
$n$ dependence of 
$\lambda(\text{A}_{1g},\text{triplet},0)$
is rather moderate,
since the $s$-wave pairing utilizes fluctuations in a larger
part of the $\mathbf{q}$ space
and it is relatively insensitive to $\mathbf{q}_{\text{max}}$.
By increasing $n$,
the magnitude of fluctuations are enhanced,
and $\lambda(\text{A}_{1g},\text{triplet},0)$ becomes large.
As shown in Fig.~\ref{figure:lambda_chi}(c),
$\lambda(\text{A}_{1g},\text{triplet},0)$ is still large for small $n$,
and the $s$-wave state may realize in a wider $n$ region
if we lower a temperature.
The eigenvalue $\lambda(\text{B}_{1g},\text{singlet},y)$
for the $d_{x^2-y^2}$-wave state also becomes large by increasing $n$,
however,
the spin susceptibility enhances more rapidly.
Here, we define a magnetic transition temperature
where $\chi^{\sigma^z}(\mathbf{q}_{\text{max}})$ becomes $50/t$.
Then, the $d_{x^2-y^2}$-wave state does not appear
within the calculated parameter region,
but it may appear by lowering temperature and/or
by adjusting parameter $J$.

\begin{figure}[t]
  \includegraphics[width=7.5cm]{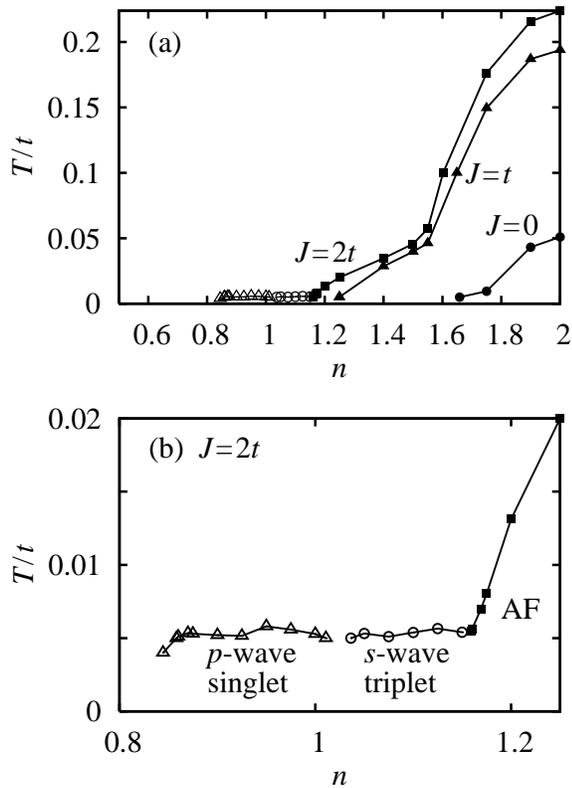}
  \caption{\label{figure:PD}
    (a) Transition temperatures for $U=6t$.
    Solid circles, triangles, and squares
    denote antiferromagnetic transition temperatures
    for $J=0$, $t$, and $2t$, respectively.
    Open circles and triangles denote
    superconducting transition temperatures for
    the $s$-wave (A$_{1g}$) spin-triplet orbital-$\tau^0$ state and for
    the $p$-wave (E$_{ u}$) spin-singlet orbital-$\tau^0$ state,
    respectively, for $J=2t$.
    (b) Transition temperatures around $n=1$ for $J=2t$.
  }
\end{figure}
Figure~\ref{figure:PD}(a) shows
the highest transition temperatures
among all the possible superconducting and antiferromagnetic ones
as functions of $n$
for $J=0$, $t$, and $2t$.
Figure~\ref{figure:PD}(b) shows the highest transition temperatures
for $J=2t$ around $n=1$.
The antiferromagnetic phase extends by increasing the Hund's rule coupling.
For $J=0$ and $t$, we cannot find any superconducting phase within
$T\ge 0.005t$.
For $J=2t$
around $n=0.9$ where $\mathbf{q}_{\text{max}}\simeq (\pi,0)$,
the $p$-wave spin-singlet orbital-$\tau^0$ state appears.
The superconducting transition temperature for the $p$-wave state
probably decreases rapidly at $n \lesssim 0.8$
as is expected from $n$ dependence of $\lambda(\text{E}_{ u},\text{singlet},0)$
shown in Fig.~\ref{figure:lambda_chi}(c),
while we cannot obtain reliable estimation of transition temperatures with
low values at present because of computational limitations, in particular,
a lattice-size limitation.
For $J=2t$ around $n\simeq 1.1$
where the antiferromagnetic transition temperature tends to zero,
the $s$-wave spin-triplet orbital-$\tau^0$ state appears.
As is expected from $n$ dependence of $\lambda(\text{A}_{1g},\text{triplet},0)$
shown in Fig.~\ref{figure:lambda_chi}(c),
the superconducting transition temperature for the $s$-wave state
depends slightly on $n$.
The $s$-wave state may appear also at $n \lesssim 0.8$
with low transition temperatures
as is expected from the behavior of $\lambda(\text{A}_{1g},\text{triplet},0)$
and $\lambda(\text{E}_{ u},\text{singlet},0)$.

We have determined the highest transition temperatures among all the possible
superconducting and antiferromagentic ones.
To discuss competition and coexistence of possible states,
we have to compare thermodynamic potentials in ordered states,
and it is one of future tasks.

\section{Pairing symmetry in other equivalent models}
\label{sec:other_models}
In this section, we discuss pairing symmetry in other models which are described
by Eq.~\eqref{eq:H}.
So far, we have assumed that the orbitals have $s$-orbital symmetry
($s$-orbital model),
but we can also consider models with different orbital symmetry
with the same Hamiltonian Eq.~\eqref{eq:H}.

The model for doubly degenerate $p_x$ and $p_y$ orbitals
with $(pp\sigma)=(pp\pi)=t$ ($p$-orbital model) is given by Eq.~\eqref{eq:H},
but the symmetry of the wavefunctions of orbitals is different from
that in the $s$-orbital model.
For example,
by the symmetry operation $x \rightarrow -x$,
the sign of the wavefunction for the $p_x$ orbital changes,
and then, the anomalous self-energy for orbital-antiparallel
(orbital-$\tau^z$ and -$\tau^0$)
states transforms to that transformed in the $s$-orbital model
multiplied by $(-1)$ under this symmetry operation.
Thus, the symmetry of the anomalous self-energy in the $p$-orbital model
is different from that in the $s$-orbital model.
Note that
the model with $d_{zx}$ and  $d_{yz}$ orbitals ($t_{2g}$ orbitals)
with $(dd\pi)=(dd\delta)=t$
is equivalent to the $p$-orbital model
including the symmetry of orbitals within the square lattice.
The model with degenerate $e_g$ orbitals with $(dd\sigma)=(dd\delta)=t$
($e_g$-orbital model) is also the same model as the $s$-orbital model
except for orbital symmetry.

\begin{table}
  \caption{\label{table:SC_sym}
    Equivalent pairing symmetry
    among $s$-orbital, $p$-orbital, and $e_g$-orbital models.
  }
  \begin{ruledtabular}
    \begin{tabular}{ccccccc}
      orbital $\tau^x$ & $s$-orbital &
      A$_{1g}$ & A$_{2g}$ & B$_{1g}$ & B$_{2g}$ & E$_{u}$ \\
                 & $p$-orbital &
      B$_{1g}$ & B$_{2g}$ & A$_{1g}$ & A$_{2g}$ & E$_{u}$ \\
                 & $e_g$-orbital &
      A$_{1g}$ & A$_{2g}$ & B$_{1g}$ & B$_{2g}$ & E$_{u}$ \\
      \hline
      
      orbital $\tau^y$ & $s$-orbital &
      A$_{1g}$ & A$_{2g}$ & B$_{1g}$ & B$_{2g}$ & E$_{u}$ \\
                 & $p$-orbital &
      A$_{1g}$ & A$_{2g}$ & B$_{1g}$ & B$_{2g}$ & E$_{u}$ \\
                 & $e_g$-orbital &
      A$_{1g}$ & A$_{2g}$ & B$_{1g}$ & B$_{2g}$ & E$_{u}$ \\
      \hline
      
      orbital $\tau^z$ & $s$-orbital &
      A$_{1g}$ & A$_{2g}$ & B$_{1g}$ & B$_{2g}$ & E$_{u}$ \\
                 & $p$-orbital &
      B$_{2g}$ & B$_{1g}$ & A$_{2g}$ & A$_{1g}$ & E$_{u}$ \\
                 & $e_g$-orbital &
      B$_{1g}$ & B$_{2g}$ & A$_{1g}$ & A$_{2g}$ & E$_{u}$ \\
      \hline
      
      orbital $\tau^0$ & $s$-orbital &
      A$_{1g}$ & A$_{2g}$ & B$_{1g}$ & B$_{2g}$ & E$_{u}$ \\
                 & $p$-orbital &
      A$_{2g}$ & A$_{1g}$ & B$_{2g}$ & B$_{1g}$ & E$_{u}$ \\
                 & $e_g$-orbital &
      B$_{1g}$ & B$_{2g}$ & A$_{1g}$ & A$_{2g}$ & E$_{u}$ \\
     \end{tabular}
  \end{ruledtabular}
\end{table}
In Table~\ref{table:SC_sym}, we show equivalent pairing symmetry
among $s$-orbital, $p$-orbital, and $e_g$-orbital models, e.g.,
the orbital-$\tau^0$ state with A$_{2g}$ symmetry ($g$-wave) in the $p$-orbital model
has the same transition temperature as
the orbital-$\tau^0$ state with A$_{1g}$ symmetry in the $s$-orbital model.
However, the wavenumber dependence of the anomalous self-energy is the same
among the equivalent states.
From the calculated results for the $s$-orbital model,
we find that the $g$-wave spin-triplet orbital-$\tau^0$ state appears
in the $p$-orbital model,
and
the $d_{x^2-y^2}$-wave spin-triplet orbital-$\tau^0$ state appears
in the $e_g$-orbital model.
The $p$-wave spin-singlet orbital-$\tau^0$ state appears
both in the $p$-orbital and the $e_g$-orbital models.


Here we discuss another model for $s$ orbitals with two Fermi-surfaces.
Under the transformation
$c_{i2\sigma} \rightarrow e^{i\mathbf{Q}\cdot \mathbf{r}_i} c_{i2\sigma}$
with $\mathbf{Q}=(\pi,\pi)$,
or equivalently $c_{\mathbf{k}2\sigma} \rightarrow c_{\mathbf{k}+\mathbf{Q}2\sigma}$,
the dispersion in Eq.~\eqref{eq:H} changes as
$\epsilon_{\mathbf{k}2} \rightarrow
\epsilon_{\mathbf{k}+\mathbf{Q}2}=-\epsilon_{\mathbf{k}2}$,
while the onsite terms in Eq.~\eqref{eq:H} are invariant.
Thus, the $s$-orbital model changes to the model described by
Eq.~\eqref{eq:H} with the dispersion
$\epsilon_{\mathbf{k}1}=-\epsilon_{\mathbf{k}2}=2t(\cos k_x+\cos k_y)$
by this transformation.

\begin{figure}[t]
  \includegraphics[width=7.5cm]{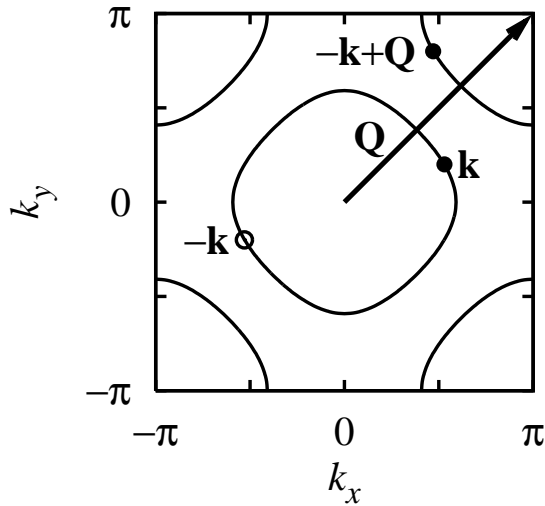}
  \caption{\label{figure:FS}
    Fermi surfaces (solid curves) for the dispersion
    $\epsilon_{\mathbf{k}1}=-\epsilon_{\mathbf{k}2}=2t(\cos k_x+\cos k_y)$
    for $n=1$.
    A pair carrying the total momentum $\mathbf{Q}$
    is composed of electrons denoted by closed circles
    on the different Fermi surfaces.
  }
\end{figure}

In such a two-Fermi-surface system,
we expect superconducting states with the finite total momentum $\mathbf{Q}$
like the FFLO state by the following reason.
In ordinary cases,
such a superconducting state with a finite total momentum $\mathbf{q}$
is hard to be realized,
since the electron on the Fermi surface with the wavevector $\mathbf{k}$
has to pair with the electron with $-\mathbf{k}+\mathbf{q}$,
but it is off the Fermi surface
unless $\mathbf{k}$ satisfies conditions depending on $\mathbf{q}$ and
the structure of the Fermi surface.
On the other hand in the model with two Fermi-surfaces,
when the electron with $\mathbf{k}$ is on a Fermi surface,
the electron with $-\mathbf{k}+\mathbf{Q}$
is always on the other Fermi surface as shown in Fig.~\ref{figure:FS}.
Thus, superconducting states with the total momentum $\mathbf{Q}$
is naturally expected in the model with two Fermi-surfaces,
where $\mathbf{Q}$ connects centers of these Fermi surfaces.

Indeed, under the transformation
$c_{\mathbf{k}2\sigma} \rightarrow c_{\mathbf{k}+\mathbf{Q}2\sigma}$,
an anomalous Green's function for a zero-total-momentum state changes as
\begin{equation}
  \begin{split}
    F_{1 \sigma; 2 \sigma^{\prime}}(\mathbf{k},\tau)
    &=-
    \langle \text{T}_{\tau}
    c_{\mathbf{k} 1 \sigma}(\tau)
    c_{-\mathbf{k} 2 \sigma^{\prime}}
    \rangle
    \\
    \rightarrow
    F_{1 \sigma; 2 \sigma^{\prime}}(\mathbf{k},\tau;\mathbf{Q})
    &=-
    \langle \text{T}_{\tau}
    c_{\mathbf{k} 1 \sigma}(\tau)
    c_{-\mathbf{k}+\mathbf{Q} 2 \sigma^{\prime}}
    \rangle,
  \end{split}
\end{equation}
where $F_{1 \sigma; 2 \sigma^{\prime}}(\mathbf{k},\tau;\mathbf{Q})$
denotes the anomalous Green's function
for a pairing state with the total momentum $\mathbf{Q}$.
On the other hand,
$F_{1 \sigma; 1 \sigma^{\prime}}(\mathbf{k},\tau)$ and
$F_{2 \sigma; 2 \sigma^{\prime}}(\mathbf{k},\tau)$ change to
$F_{1 \sigma; 1 \sigma^{\prime}}(\mathbf{k},\tau)$ and
$F_{2 \sigma; 2 \sigma^{\prime}}(\mathbf{k}+\mathbf{Q},\tau)$, respectively,
i.e.,
these anomalous Green's functions are corresponding to
pairing states with zero total momentum.
Thus, by this transformation,
the orbital-antiparallel superconducting states
(orbital-$\tau^z$ and -$\tau^0$ states)
change to those with the finite total momentum $\mathbf{Q}$,
while the orbital-parallel states
(orbital-$\tau^x$ and -$\tau^y$ states)
remain those with zero total momentum.
Concerning the susceptibilities,
$\chi^{\text{A}}(\mathbf{q})$ transforms to
$\chi^{\text{A}}(\mathbf{q}+\mathbf{Q})$
for $\text{A}=\tau^x$, $\tau^y$, $\tau^x \sigma^{\nu}$, and $\tau^y \sigma^{\nu}$,
while the other susceptibilities are unchanged.
Then, the antiferromagnetic transition temperatures are unchanged.

Thus, from the calculated results for the $s$-orbital model,
we obtain $s$-wave spin-triplet and $p$-wave spin-singlet states
with the finite total momentum $\mathbf{Q}$, even without a magnetic field,
in the model with two Fermi-surfaces
which locate around $\mathbf{k}=(0,0)$ and $\mathbf{k}=\mathbf{Q}$.
At temperatures lower than the transition temperatures,
a superconducting state may coexist with
an antiferromagnetic (spin density wave) state.
A superconducting state with the total momentum $\mathbf{Q}$
may be stabilized further by coexisting with
a spin density wave with $\mathbf{Q}$,~\cite{Psaltakis,Gulacsi}
and it is an interesting future problem.

Note also that a model for $p_x$ and $p_y$ orbitals
with $-(pp \sigma)=(pp \pi)=t$,
in which Fermi surfaces locate around
$\mathbf{k}=\mathbf{Q}_1=(\pi,0)$ and
$\mathbf{k}=\mathbf{Q}_2=(0,\pi)$,
is equivalent to the $p$-orbital model with $(pp \sigma)=(pp \pi)$.
We can show this equivalence by applying the transformation
$c_{i 1 \sigma} \rightarrow e^{i \mathbf{Q}_1 \cdot \mathbf{r}_i} c_{i 1 \sigma}$ and 
$c_{i 2 \sigma} \rightarrow e^{i \mathbf{Q}_2 \cdot \mathbf{r}_i} c_{i 2 \sigma}$.
Thus, we also obtain superconducting states
with the total momentum $\mathbf{Q}=\mathbf{Q}_1+\mathbf{Q}_2$
in the model with $-(pp \sigma)=(pp \pi)=t$.
This conclusion also applies to a model for $t_{zx}$ and $t_{yz}$ orbitals
with $(dd \pi)=-(dd \delta)=t$.

\section{Summary}
\label{sec:summary}
We have studied a two-orbital Hubbard model on a square lattice.
In this model, we have considered orbitals with $s$-orbital symmetry.
In such a multi-orbital system,
even-parity spin-triplet states and odd-parity spin-singlet states
are allowed even within states with even frequency dependence.
Indeed, we have found that
the $s$-wave spin-triplet orbital-antisymmetric state
and
the $p$-wave spin-singlet orbital-antisymmetric state naturally
appear within a fluctuation exchange approximation.

Tendencies toward a $s$-wave spin-triplet state
have been found also in similar two-orbital models
on infinite dimensional lattices
by using a dynamical mean-field theory.~\cite{Han,Sakai}
This fact indicates that
tendencies toward $s$-wave spin-triplet states
are common to two-orbital models
irrespective of dimensionality.

We have also discussed pairing symmetry in
equivalent models which can be described by the two-orbital Hubbard model.
The equivalent models with orbital symmetry other than $s$-orbital one
require special conditions, e.g.,
$(pp\sigma)=(pp\pi)$ for the $p$-orbital model,
and may be unrealistic.
However, we believe that these models provide
an interesting view point to discuss and determine pairing symmetry
in multi-orbital systems.
Note also that, in a realistic model
such as a model for $p$ orbitals with $(pp\sigma) \ne (pp\pi)$,
orbital states of a pair mix in general, and then,
$\mathbf{k}$ dependence of the anomalous self-energy
is not that of an irreducible representation of a system.

Finally, we have discussed models with two Fermi-surfaces.
In such multi-Fermi-surface systems,
we naturally expect a superconducting state with a finite total momentum
like the FFLO state.
Indeed, we have shown that 
the superconducting states found in the $s$-orbital model
are corresponding to pairing states with a finite total momentum,
even without a magnetic field,
in a model with two Fermi-surfaces.
Although such a system with multi-Fermi-surfaces with the same structure
is unrealistic,
we expect that such exotic states will be realized
also in realistic models with multi-Fermi-surfaces whose centers locate around
different $\mathbf{k}$ points with similar,
but not exactly the same, structures.

\section*{Acknowledgments}
The author thanks T. Hotta for
a critical reading of the manuscript and useful comments.
The author also thanks H. Onishi for helpful comments.
The author is supported by Grants-in-Aid for Scientific Research
in Priority Area ``Skutterudites''
and for Young Scientists
from the Ministry of Education, Culture, Sports, Science,
and Technology of Japan.


\end{document}